\documentclass{article}
\usepackage{ijcai22}

\usepackage{times}
\usepackage{soul}
\usepackage{url}
\usepackage[hidelinks]{hyperref}
\usepackage[utf8]{inputenc}
\usepackage[small]{caption}
\usepackage{graphicx}
\usepackage{amsmath,amscd,amsbsy,amssymb,amsfonts,latexsym,url,bm,amsthm}
\usepackage{booktabs,subfigure,longtable,threeparttable,multirow,longtable,textcomp,dsfont}
\usepackage{algorithm}
\usepackage{algorithmic}
\newtheorem{theorem}{Theorem}
\newtheorem{proposition}{Proposition}

\title{Trading Hard Negatives and True Negatives: \\
A Debiased Contrastive Collaborative Filtering Approach}

\author{
Chenxiao Yang$^1$
\and
Qitian Wu$^1$\and
Jipeng Jin$^1$\and
Xiaofeng Gao$^1$\footnote{This work was supported by the National Key R\&D Program of China [2020YFB1707903]; the National Natural Science Foundation of China [61872238, 61972254], the Shanghai Municipal Science and Technology Major Project [2021SHZDZX0102], and the Tencent Rhino-Bird Renewed Research Program. Xiaofeng Gao is the corresponding author.}\and
Junwei Pan$^2$\And
Guihai Chen$^1$
\affiliations
$^1$MoE Key Lab of Artificial Intelligence, Department of Computer Science and Engineering, \\Shanghai Jiao Tong University\\
$^2$Tencent Inc.
\emails
\{chr26195, echo740, jinjipeng,gaoxiaofeng\}@sjtu.edu.cn,
jonaspan@tencent.com,
gchen@cs.sjtu.edu.cn
}

\begin{document}

\maketitle

\begin{abstract}
Collaborative filtering (CF), as a standard method for recommendation with implicit feedback, tackles a semi-supervised learning problem where most interaction data are unobserved. Such a nature makes existing approaches highly rely on mining negatives for providing correct training signals. However, mining proper negatives is not a free lunch, encountering with a tricky trade-off between mining informative hard negatives and avoiding false ones. We devise a new approach named as Hardness-Aware Debiased Contrastive Collaborative Filtering (HDCCF) to resolve the dilemma. It could sufficiently explore hard negatives from two-fold aspects: 1) adaptively sharpening the gradients of harder instances through a set-wise objective, and 2) implicitly leveraging item/user frequency information with a new sampling strategy. To circumvent false negatives, we develop a principled approach to improve the reliability of negative instances and prove that the objective is an unbiased estimation of sampling from the true negative distribution. Extensive experiments demonstrate the superiority of the proposed model over existing CF models and hard negative mining methods. 
\end{abstract}

\section{Introduction}
Collaborative Filtering (CF)~\cite{pan2008one} is a standard approach to deal with implicit feedback (e.g., click, watch, purchase, etc.) in recommender systems, wherein observed user-item interactions are assigned with positive labels, and the rest are unlabeled. A common practice in CF methods is to uniformly draw negative instances from the unlabeled portion, a.k.a. \emph{negative sampling}~\cite{chen2017sampling}, and then use both positive and negative instances for training, as has been adopted by existing \emph{point-wise}~\cite{mnih2008probabilistic} or \emph{pair-wise}~\cite{rendle2009bpr} approaches. However, this CF paradigm is considered as insufficient to provide informative and reliable training signals. Hence, enormous efforts have been made for improving the quality of negative instances for CF-based recommendation. 

Particularly, \emph{hard negative mining} has shown to be an effective approach, which aims to exploit negative user-item pairs whose embeddings are close yet expected to be far apart~\cite{wu2017sampling,park2019adversarial}, as a means to provide informative training signals. A line of works~\cite{rendle2014improving,chen2017sampling} fallen into this category attempt to replace the uniform negative sampling distribution by some predefined surrogates, based on certain prior knowledge such as that more frequent items constitutes better negatives~\cite{wu2019noise,chen2017sampling}. In contrast, another line of works~\cite{rendle2014improving,park2019adversarial} seek to adaptively mine negatives by carefully examining relevance score of user-item pairs, which are generally more effective but often require sophisticated training techniques such as generative adversarial network~\cite{park2019adversarial}, reinforcement learning~\cite{ding2019reinforced} and bi-level optimization~\cite{shu2019meta}. This leads to the \textbf{first} trade-off between efficiency and effectiveness.

On the other hand, the soundness of these works resides on a problematic assumption that ``all unlabeled interactions are true negative instances", which is against the actual setting where unlabeled user-item pairs may potentially become positive instances, once the item is exposed to the user. These instances are termed as \emph{false negatives}. The incorporation of false negatives would provide erroneous supervised signals for training and seriously degrade the performance~\cite{hernandez2014probabilistic}. 
While it may sound attractive to identify and remove these  instances, it is challenging to distinguish hard negatives and false negatives, given that both of them have large relevance scores in appearance and auxiliary information is often not available. Few works attempt to address this issue, especially in the context of negative mining for CF. This presents the \textbf{second} trade-off between informative negatives and reliable negatives. 

Towards navigating these trade-offs, we propose a new framework named as Hardness-Aware Debiased Contrastive Collaborative Filtering (HDCCF). 
Specifically, a contrastive loss function is devised in place of conventional point-wise and pair-wise objectives, which will be shown by gradient analysis that can automatically and adaptively concentrate on optimizing hard negatives by contrasting with peers, notably, without relying on complex training tricks. 
We also devise an efficient sampling strategy that implicitly explores negative instances by incorporating item frequency information, without actually conducting negative sampling. 
On top of the new hardness-aware objective, we further propose a principled method to eliminate the risk of false negatives. Needless to explicitly distinguish hard and false negatives, this is achieved by directly debiasing the objective, such that its expectation is strictly equivalent to the ideal loss function that resembles sampling under true negative distribution. 

There are also several additional novel designs in our framework: 1) It considers both negative users and items to avoid the case where all negative items are relatively discriminative for a specific user, and vice versa; 2) Two auxiliary contrastive losses are introduced to model user-user and item-item relationship, which could help to obtain more meaningful user and item representations; 3) A neural modulated mechanism is designed that takes a user's diverse preference on different items into account in the loss function. We validate the effectiveness of HDCCF by comparison experiments and ablation studies. The results demonstrate the superiority of HDCCF as well as the effectiveness of its components. 

\begin{figure*}[t]
	\centering
	\includegraphics[width=0.88\textwidth,angle=0]{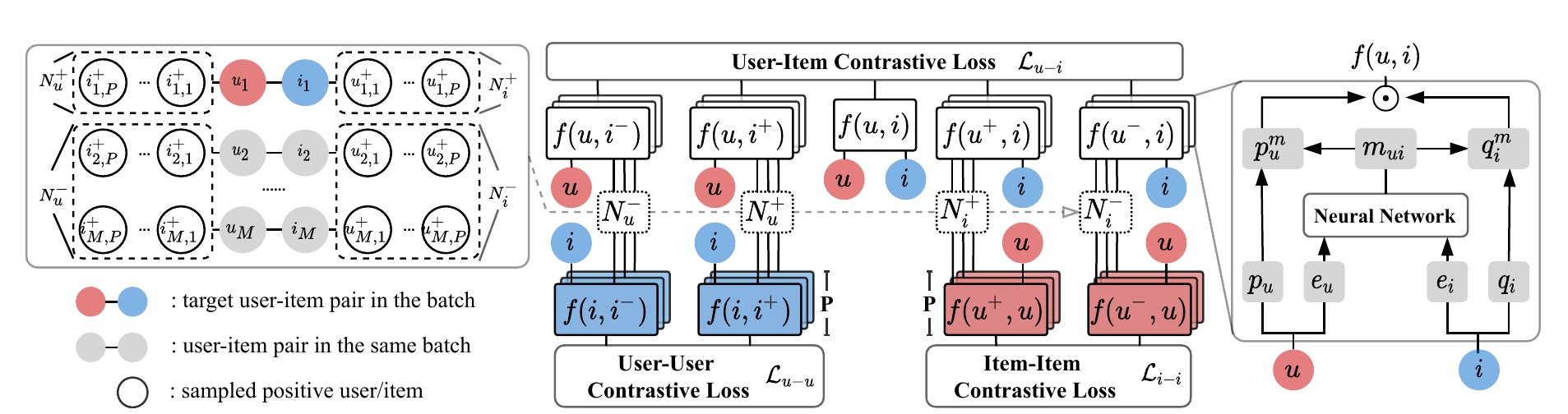}
	\caption{Overview of Hardness-Aware Debiased Contrastive Collaborative Filtering (HDCCF) framework.}
	\label{fig:frame}
\end{figure*} 

\section{Proposed Model}
\paragraph{Notations.} Let $\mathcal{U}$ and $\mathcal{I}$ denote a set of users and items, $\mathcal{I}_u$ (resp. $\mathcal{I}'_u$) denote a set of items that user $u$ has (resp. has not) interacted with, $\mathcal{U}_i$ (resp. $\mathcal{U}'_i$) denote a set of users that item $i$ have (resp. not) interacted with. Observed interaction data are represented by a set of user-item pairs $\mathcal{D} = \{(u,i)\}$. Unobserved user-item pairs are denoted as $\mathcal{D}' = \mathcal{U} \times \mathcal{I} - \mathcal{D}$. For user $u\in \mathcal{U}$, our goal is to recommend a fix-sized set of ordered items $\mathcal{X}_u \subset \mathcal{I}'_u$.

\subsection{Hardness-Aware Contrastive CF}

For an observed instance $(u,i) \in \mathcal{D}$, we uniformly sample $S$ negative items (resp. users) that have no observed interaction with user $u$ (resp. item $i$), denoted as $\mathcal{N}_u^- \subset \mathcal{I}'_u$ (resp. $\mathcal{N}_i^- \subset \mathcal{U}'_i$), where $S$ is the \emph{negative sampling number}. Then, the user-item contrastive loss $\mathcal{L}_{u-i}$ is defined as
\begin{equation}\label{eqn-uiloss}
\resizebox{.91\linewidth}{!}{$
\displaystyle
    \mathcal{L}_{u-i} = - \sum_{(u,i)\in \mathcal{D}} \log \frac{e^ {f(u,i)/\tau}}{F^i(\mathcal{N}_i^-) + F^u(\mathcal{N}_u^-) + e^ {f(u,i)/\tau}},
$}
\end{equation} 
where $f$ : $(u,i) \rightarrow \mathbb R$ is a similarity measure which outputs the relevance score of user $u$ and item $i$, and $\tau \in \mathbb{R}^+$ is a scalar temperature parameter (omitted in the following for brevity). The $F^u$ (resp. $F^i$) in Eqn.~\eqref{eqn-uiloss} is called \emph{negative score} for user $u$ (resp. item $i$), and is formulated as 
\begin{equation}
\resizebox{.91\linewidth}{!}{$
\displaystyle
    F^u(\mathcal{N}_u^-) = \sum\limits_{i^- \in \mathcal{N}_u^-} e^{f(u,i^-)}, \quad F^i(\mathcal{N}_i^-) = \sum\limits_{u^- \in \mathcal{N}_i^-}  e^{f(u^-,i)},
    $}
\end{equation} 
where both $(u,i^-)$ and $(u^-,i)$ are unobserved (i.e., unlabeled) instances.

\paragraph{Hardness-Aware Property by Gradient Analysis.}
To explain the efficacy of the above contrastive loss formulation for hard negative mining, we peer into its gradients with respect to observed and unobserved instances for analysis. Denote the probability of an unobserved instance $(u',i')$ being recognized as positive by
\begin{equation}
\begin{split}
    P(u',i') &= \frac{e^ {f(u',i')/\tau}}{F^i(\mathcal{N}_i^-) + F^u(\mathcal{N}_u^-) + e^ {f(u,i)/\tau}},\\
    \text{where}\quad (u'&,i') \in \{(u,i^-)\}_{i^- \in \mathcal{N}_u^-} \cup \{(u^-,i)\}_{u^- \in \mathcal{N}_i^-}
\end{split}
\end{equation}
Then, the gradients with respect to the relevance score of observed and unobserved instances are computed as
\begin{equation}
\resizebox{.91\linewidth}{!}{$
\displaystyle
    \frac{\partial \mathcal L_{u-i}}{\partial f(u,i)} = -\frac{1}{\tau} \sum_{(u',i')} P(u',i'),\;
    \frac{\partial \mathcal L_{u-i}}{\partial f(u',i')} = \frac{1}{\tau} P(u',i'),
    $}
\end{equation}
These equations reveal two properties~\cite{wang2021understanding}:
\begin{enumerate}
    \item For each individual unobserved instance $(u',i')$, the gradient is proportional to $P(u',i')$ and thus is also proportional to $e^{f(u',i')/\tau}$;
    \item The gradient for observed instance is equal to the sum of gradients for all unobserved instances.
\end{enumerate}
These properties have several implications in the context of hard negative mining in CF. 
\textbf{First}, according to the first property, a \emph{harder} negative instance with larger relevance score has larger magnitude of gradients, which indicates the loss function could automatically concentrates on optimizing harder negative instances. The hardness level for each negative instance is adaptively updated for each iteration, and could be controlled by tuning temperature $\tau$. \textbf{Second}, the gradient is re-scaled by the sum of relevance scores of peer negative instances, which indicates the hardness for each negative instance is relative to the hardness of peer negatives in the loss function, distinguishing us from pair-wise loss functions. \textbf{Third}, according to the second property, the gradients of negative instances, whose sum is determined by ${\partial \mathcal L_{u-i}}/{\partial f(u,i)}$, are distributed over each negative instance, and thus are not sensitive to label noise, which is known as a limitation of BPR loss. \textbf{Fourth}, by considering two types of negative instances for every $(u,i)$ in Eqn.~\eqref{eqn-uiloss} (i.e., negative items and users), we could jointly mine negative instances from two facets and avoid the case when all negative items (resp. users) are easily discriminative for an individual user (resp. item). 

\paragraph{User-User and Item-Item Relations.}
Besides modeling user-item interactions, we further extend the advantage of negative mining to neighbored users (i.e. a pair of users $(u,u')$ that have interactions with the same item) and neighbored items~\cite{sarwar2001item,kabbur2013fism} by proposing two auxiliary contrastive losses. We uniformly sample $P$ positive items (resp. users) for user $u$ (resp. item $i$), denoted as $\mathcal{N}_u^+$ (resp. $\mathcal{N}_i^+$), where $P$ is the \emph{positive neighbor sampling number}. The auxiliary loss function $\mathcal L_{u-u}$ is formulated as:
\begin{equation} \label{eqn:u-u}
\resizebox{.91\linewidth}{!}{$
\displaystyle
    \mathcal{L}_{u-u} = - \sum_{(u,i)\in \mathcal{D}} \sum_{u^+ \in \mathcal{N}_i^+} \log \frac{e^{f(u,u^+)}}{e^{f(u,u^+)} + \sum\limits_{u^- \in \mathcal{N}_i^-}e^{f(u,u^-)}},
    $}
\end{equation}
where $\mathcal{N}_u^-$ and $\mathcal{N}_i^-$ are the same as those used in Eqn.\eqref{eqn-uiloss}, and $\mathcal L_{i-i}$ could be defined in the same way. Then, the final loss function $\mathcal{L}$ is the weighted sum of three terms:
\begin{equation} \label{eqn:loss}
    \mathcal{L} = \mathcal L_{u-i} + \lambda_u \mathcal L_{u-u} + \lambda_i \mathcal L_{i-i},
\end{equation}
where $\lambda_u$ and $\lambda_i$ are weights to balance the importance for each type of relation. 

\paragraph{Neural Modulated Similarity Model.}
For the similarity function $f(u,i)$, one could use dot-product~\cite{koren2009matrix}, Euclidean distance~\cite{hsieh2017collaborative} or parameterize it with neural network~\cite{he2017neural}. Particularly in this paper, we propose to use the following neural modulated similarity model:
\begin{equation}
    f(u,i) = (\mathbf{m}_{ui} \odot \mathbf{p}_u)^\top \cdot (\mathbf{m}_{ui} \odot \mathbf{q}_{i}),
\end{equation}
where $\mathbf{p}_{u}, \mathbf{q}_{i} \in \mathbb{R}^{d}$ are user and item embeddings, and $\mathbf{m}_{ui} \in \mathbb{R}^{d}_{+}$ is a modulating vector for element-wise scaling, which is computed by
\begin{equation}
    \mathbf{m}_{ui} = \sigma \left(g\big(\mathbf{e}_i\| \mathbf{e}_u\| (\mathbf{e}_i \odot \mathbf{e}_u)\right),
\end{equation}
where $\mathbf{e}_u, \mathbf{e}_i\in \mathbb R^d$ are user and item latent factors, $\odot$ denotes Hadamard product, $\|$ denotes vector concatenation, and $g: \mathbb R^{3d} \rightarrow \mathbb R^d$ is a neural network. The key insight is that using a fixed user embedding may fail to represent one's diverse preference on distinct items in the loss function (items may also have multiple attributes that could attract a user), especially in our case where the objective incorporate more candidates of items. To mitigate this issue, our design could capture a user’s varying preferences by allowing more flexible representations, which could improve the discrimination ability of the model, empirically verified by ablation studies. 

\subsection{Sampling Strategy}\label{sampling}
The sampling approach mentioned in last subsection samples $S$ negative users/items and $P$ positive users/items for each target observed user-item pair. Suppose the batch size is $M$, there are additional $2M \times(S+P)$ (where $M\approx S$, $S\gg P$) users/items to be sampled for each iteration besides target user-item pairs, which is impractical when scaling the training. Alternatively, we adopt a sampling strategy that uses the positive users (resp. items) from other observed instances in the same mini-batch as the negative users (resp. items) for the target instance. Formally,
\begin{equation}
\resizebox{.91\linewidth}{!}{$
\displaystyle
    \mathcal{N}_{u_k}^- = \mathop{\cup}\limits_{m\in\{1,\cdots, M\}\backslash k} \mathcal{N}_{u_m}^+, \quad\mathcal{N}_{i_k}^- = \mathop{\cup}\limits_{m\in\{1,\cdots, M\}\backslash k} \mathcal{N}_{i_m}^+, 
    $}
\end{equation}
where $\mathcal{N}_{u_k}^-$ and $\mathcal{N}_{i_k}^-$ are multi-sets, which allow multiple appearances of the same item or user in the set. In this way, the negative instance number is enlarged from $S$ to $P\times (M-1)$ with lower sampling overhead (from $2M \times(S+P)$ to $2M \times P$). We also highlight that such sampling strategy is free of explicit negative sampling.

\paragraph{Frequency-Aware Properties.} To further shed lights on HDCCF's hard negative mining capability and its relation with frequency-based sampling methods, we investigate on: for a specific positive user-item interaction $(u,i)$ in a mini-batch, 1) the number of times a negative item $i'$ (resp. $u'$) appears in user-item contrastive loss, denoted as $n_{i'}^{u-i}$ (resp. $n_{u'}^{u-i}$) and 2) the number of times a negative item $i'$ (resp. $u'$) appears in auxiliary contrastive losses, denoted as $n_{i'}^{i-i}$ (resp. $n_{u'}^{u-u}$).

\begin{proposition}
The expectation of $n_{i'}^{u-i}$ is proportional to the number of times this item appears in interaction dataset $\mathcal{D}$, i.e., $n_{i'}^{u-i} \propto |\mathcal{U}_{i'}|$. This property holds true for user $u'$ in $\mathcal{L}_{u-i}$. Formally we have
\begin{equation}
\begin{split}
    \mathbb{E}_{(u,i)\sim p^o\atop i^+ \sim p^o_u}\left[n_{i'}^{u-i}\right] &= \frac{M-1}{|\mathcal D|-1} \cdot P\cdot |\mathcal{U}_{i'}|,\\
    \mathbb{E}_{(u,i)\sim p^o\atop u^+ \sim p^o_i}\left[n_{u'}^{u-i}\right] &=  \frac{M-1}{|\mathcal D|-1} \cdot P\cdot |\mathcal{I}_{u'}|,
\end{split}
\end{equation}
where the observed interaction $(u,i)$ is sampled from $\mathcal{D}$ with distribution $p^o(u,i)$, and the neighbored item $i^+$ is sampled from $\mathcal{I}_u$ with distribution $p^o_u(i^+)$.
\end{proposition}

\begin{proposition}
The expectation of $n_{i'}^{i-i}$ is also proportional to $|\mathcal{I}_{u'}|$. This property holds true for user $u'$ in user-user contrastive loss $\mathcal{L}_{u-u}$. Formally we have
\begin{equation}
\begin{split}
    \mathbb{E}_{(u,i)\sim p^o\atop i^+ \sim p^o_u}\left[n_{i'}^{i-i}\right] &= \frac{M-1}{|\mathcal D|-1} \cdot P\cdot |\mathcal{U}_{i'}|,\\
    \mathbb{E}_{(u,i)\sim p^o\atop u^+ \sim p^o_i}\left[n_{u'}^{u-u}\right] &=  \frac{M-1}{|\mathcal D|-1} \cdot P\cdot |\mathcal{I}_{u'}|,
\end{split}
\end{equation}
\end{proposition}
The proof of these propositions is shown in the appendix. As an observation, both $\mathbb{E}[n_{i'}^{u-i}]$ and $\mathbb{E}[n_{i'}^{i-i}]$ are proportional to $|\mathcal{U}_{i'}|$, which indicates that such sampling strategy is essentially frequency-aware, which enforces the loss function to concentrate on more frequent (popular) items. Since popular items are treated as harder negative instances~\cite{chen2017sampling,wu2019noise}, such sampling strategy implicitly agrees with the negative mining efficacy of HDCCF.

\subsection{Debiased Contrastive Loss} \label{debias}
As mentioned before, a user-item pair $(u,i^-)$ that we regard as a negative instance is potentially a positive interaction (i.e., \emph{false negative instance}). The existence of false negative instance could introduce bias in the training signals, and hence may cause sub-optimal results. Particularly, in our case, there are two types of false negative instances: 1) The user-item pair is an observed instance, i.e., $(u, i^-) \in \mathcal{D}$; 2) Though $(u, i^-)$ is unobserved, the interaction will occur once $i^-$ is exposed to $u$. In a similar spirit with~\cite{robinson2021contrastive} that considers a simpler case (without the first type of false negatives and user-user/item-item losses) in the general contrastive learning setting, we propose to eliminate the effects of false negatives by first formulating the expected loss function and then devising unbiased versions of Eqn.~\eqref{eqn-uiloss} and Eqn.~\eqref{eqn:u-u} without violating their hardness-aware properties.

\paragraph{Formulation of Expected Loss Function.} Our analysis mainly focus on the user side for brevity, while the same also applies to the item side. Given an item $i$, suppose a negative user $u^-$ is drawn from $\mathcal{U}'_i$ with a \emph{negative sampling distribution} $p_i(u^-)$, i.e., a uniform distribution. Drawing from this distribution may either yield a false negative instance or a real negative instance. Suppose their probabilities are $\omega_u^+$ and $\omega_u^-$ (i.e., $1-\omega_u^+$) respectively. To investigate on the formulation of expected loss function, we denote $p^+_i(u^-)$ (resp. $p^-_i(u^-)$) as the sampling distribution for false (resp. real) negative instance, which are unknown for us. The marginalization of the negative sampling distribution induces a decomposition form $p_i(u') = \omega_u^+ \cdot p_i^+(u') + \omega_u^- \cdot p_i^-(u')$. Reorganizing it yields the following expression for real negative sampling distribution
\begin{equation} \label{eqn:decompose}
p_i^-(u')=\left\{
\begin{aligned}
&0  &,  (u',i)\in \mathcal{D}, \\
\frac{p_i(u')}{\omega_u^-} &- \frac{\omega_u^+ \cdot p_i^+(u')}{\omega_u^-}  &,  (u',i)\notin \mathcal{D}.
\end{aligned}
\right.
\end{equation}
Equipped with these notations, we can formulate the ideal optimization objective for $\mathcal{L}_{u-i}$ as
\begin{equation} \label{eqn:ideal}
\resizebox{.89\linewidth}{!}{$
\displaystyle
\begin{split}
    &\mathcal{L}_{u-i}^{ideal} = -\mathop{\mathbb{E}}\limits_{(u,i)\sim p^o} \\
    &\left[ \frac{e^{f(u,i)}}{e^{f(u,i)} + Q\mathop{\mathbb{E}}\limits_{u^- \sim p_i^-}[e^{f(u^-, i)}] + Q\mathop{\mathbb{E}}\limits_{i^- \sim p_u^-}[e^{f(u, i^-)}]} \right],
\end{split}
$}
\end{equation}
where $Q$ is constant to facilitate the analysis. By comparison between the original formulation of optimization objective in Eqn.~\eqref{eqn-uiloss} and the expected objective in Eqn.~\eqref{eqn:ideal}, we immediately notice the bias essentially stems from the underlying negative sampling distribution. Eliminating the effects of false negatives boils down to approximating the ideal optimization objective using biased observations in datasets. This is challenging due to the existence of two types of false negatives as stated before, and the intractability of the real negative sampling distribution.

\paragraph{Modification of Eqn.\eqref{eqn-uiloss}}
Toward eliminating the effects of false negatives, we proceed to modify both formulations of Eqn.~\eqref{eqn-uiloss} and Eqn.~\eqref{eqn:u-u} such that they agree with the ideal optimization objective. Specifically, the debiased user-item contrastive loss can be formulated as:
\begin{equation} \label{unbiased_ui}
\resizebox{.91\linewidth}{!}{$
\displaystyle
\begin{split}
    &\tilde{\mathcal{L}}_{u-i} = \\
    &-\sum_{(u,i)\in \mathcal{D}} \log \frac{e^{f(u,i)}}{\tilde{F}^u(\mathcal{N}_i^-, \mathcal{N}_i^+) + \tilde{F}^i(\mathcal{N}_u^-, \mathcal{N}_u^+) + e^{f(u,i)}},
\end{split}
$}
\end{equation} 
where $\tilde{F}^u(\mathcal{N}_i^-, \mathcal{N}_i^+)$ and $\tilde{F}^i(\mathcal{N}_u^-, \mathcal{N}_u^+)$ are debiased negative scores for negative users and negative items respectively, and the former one is defined as:
\begin{equation}
\resizebox{.93\linewidth}{!}{$
\displaystyle
\sum\limits_{u^-\in \mathcal{N}_i^-} \pi^u_0(u^-,i)\cdot e^{f(u^-, i)} - \sum\limits_{u^+\in \mathcal{N}_i^+ \cup \{u\}} \pi^u_1(u^+,i)\cdot e^{f(u^+, i)},
$}
\end{equation}
where $\pi^u_0(u^-,i), \pi^u_1(u^+,i) \in \mathbb{R}$ are constants w.r.t. $|\mathcal{N}_i^-|$, $|\mathcal{N}_i^+|$ and $\omega_u^+$. Their exact formulations will be given in the appendix. By replacing $u$ by $i$, we can get $\tilde{F}^i(u, i, \mathcal{N}_i^-, \mathcal{N}_i^+)$ in the same way. 

\begin{theorem}
Equation.~\eqref{unbiased_ui} is an unbiased estimation of the ideal user-item contrastive loss where negative instances are drawn from the real negative distributions $p_i^-$ and $p_u^-$.
\end{theorem}
The proof is shown in appendix. To prevent negative values in the logarithm, we can constrain the negative scores to be greater than its theoretical lower bound in practice
\begin{equation}
\begin{split}
    \tilde{F}^u(\mathcal{N}_i^-, \mathcal{N}_i^+) &\gets \max \big\{\tilde{F}^u(\mathcal{N}_i^-, \mathcal{N}_i^+),|\mathcal{N}_i^-|e^{1/\tau}\big\},\\
    \tilde{F}^i(\mathcal{N}_u^-, \mathcal{N}_u^+) &\gets \max\big\{\tilde{F}^i(\mathcal{N}_u^-, \mathcal{N}_u^+),|\mathcal{N}_u^-|e^{1/\tau}\big\}.
\end{split}
\end{equation}

\paragraph{Modification of Eqn.\eqref{eqn:u-u}}
In the similar spirit, we develop unbiased formulations for user-user and item-item contrastive losses $\mathcal{L}_{u-u}$ and $\mathcal{L}_{i-i}$ in the following forms
\begin{equation} \label{unbiased_uu}
\resizebox{.91\linewidth}{!}{$
\displaystyle
\begin{split}
    &\tilde{\mathcal{L}}_{u-u} = -\\
    &\sum_{(u,i)\in \mathcal{D}}\sum_{u^+ \in \mathcal{N}_i^+}
    \log \frac{e^{f(u,u^+)}}{e^{f(u,u^+)} + \sum\limits_{u^- \in \mathcal{N}_i^-}\pi^{u-u}(u,u^-)\cdot e^{f(u,u^-)}},\\
\end{split}
$}
\end{equation}
$\mathcal{D}^{u-u}$ is a set of observed neighbored users. By replacing $u$ by $i$, we can get $\Tilde{\mathcal{L}}_{i-i}$ in the same way.

\begin{table*}[t]
	\centering
\scalebox{0.85}{\setlength{\tabcolsep}{3mm}{
\begin{tabular}{c|c|ccccccc|c|c}
\toprule[1pt]
\specialrule{0em}{1pt}{1pt}
Datasets                        & Metrics & PMF    & BPR & SVD++     & NeuMF    & ENMF  & IRGAN  & SD-GAR       & HDCCF & \textit{Imp.}           \\ \specialrule{0em}{1pt}{1pt} \hline \specialrule{0em}{1pt}{1pt}
\multirow{4}{*}{ML-1M}   & HR@10   & 0.7109 & 0.7162  & 0.7230   & 0.6991 & 0.7273 & 0.7205 & \underline{ 0.7323}       & \textbf{0.7596} & 3.72\% \\
                                & NDCG@10 & 0.4396 & 0.4435  & 0.4496  & 0.4283 & 0.5193       & 0.4705 & \underline{ 0.5320}  & \textbf{0.5788} & 8.80\% \\
                                & NDCG@50 & 0.5048 & 0.5069  & 0.5132  & 0.4936  & 0.5731       & 0.5310  & \underline{ 0.5842} & \textbf{0.6233} & 6.69\% \\ \specialrule{0em}{1pt}{1pt} \hline \specialrule{0em}{1pt}{1pt}
\multirow{4}{*}{Yelp}           & HR@10   & 0.3048 & 0.3093  & 0.3268  & 0.3230   & \underline{ 0.3710} & 0.3197 & 0.3459       & \textbf{0.3911} & 5.41\% \\
                                & NDCG@10 & 0.1609 & 0.1644  & 0.1830   & 0.1835 & \underline{ 0.2212} & 0.1776 & 0.1929       & \textbf{0.2413} & 9.09\% \\
                                & NDCG@50 & 0.2181 & 0.2209  & 0.2413 & 0.2453  & \underline{ 0.2763} & 0.2470  & 0.2684       & \textbf{0.2983} & 7.96\% \\ \specialrule{0em}{1pt}{1pt} \hline \specialrule{0em}{1pt}{1pt}
\multirow{4}{*}{Gowalla}        & HR@10   & 0.7341 & 0.7402  & 0.7303  & 0.7456  & 0.7895       & 0.7721 & \underline{ 0.8060} & \textbf{0.8311} & 3.11\% \\
                                & NDCG@10 & 0.5353 & 0.5411  & 0.5292  & 0.5498 & 0.6124       & 0.5894 & \underline{ 0.6321} & \textbf{0.6633} & 4.94\% \\
                                & NDCG@50 & 0.5693 & 0.5736  & 0.5672  & 0.5777  & 0.6390       & 0.6102 & \underline{ 0.6631} & \textbf{0.6895} & 3.98\% \\  \specialrule{0em}{1pt}{1pt} \bottomrule[1pt]
\end{tabular}}}
\vspace{-5pt}
\caption{Experiment results of HDCCF and competitors. The bold value marks the best one in one row, while the underlined value corresponds to the best one among all the baselines. Improvements are statistically significant with $p < 0.01$.} \label{tab:comparison}
\vspace{-10pt}
\end{table*}

\section{Related Works}
Negative mining plays an important role in CF approaches, with significant influences on the recommendation performance~\cite{chen2017sampling,rendle2009bpr}. A high-quality negative instance should satisfy: 1) It should be hard for model to discriminate, so as to provide useful information for training~\cite{wu2017sampling}; 2) It should be a reliable negative instance sampled from the distribution of true negatives, rather than those erroneously recognized as negatives~\cite{hernandez2014probabilistic}. Most existing works on hard negative mining~\cite{ding2019reinforced,ding2018improved,park2019adversarial} either learns the sampling distribution with a separate model, or generates negative instances with adversarial training. Despite of the promising results, they often require complex designs, architectures, or side information such as user's ``view" behavior which is not always available~\cite{ding2018improved}. Besides, the risk of false negative instances are overlooked in these works. Deviating from those works, we devise a hardness-aware loss function that can automatically detect hard and reliable negatives. 

Existing optimization objectives in CF approaches could be roughly categorized into point-wise~\cite{mnih2007probabilistic}, pair-wise~\cite{rendle2009bpr} and list-wise~\cite{wu2018sql}. 
The most relevant approaches are based on list-wise, which also consider multiple instances in the loss function. However, state-of-the-art list-wise approach~\cite{wu2018sql} based on a permutation probability only optimize the upper bound rather than the original negative log-likelihood. Another related work in CF~\cite{chen2020efficient} that is also free of negative sampling essentially relies on a pair-wise objective with predefined frequency-based weight. 
Collaborative filtering has been extensively studied in other recommendation situations, e.g., social recommendation~\cite{ma2008sorec,wu2019dual}, sequential recommendation~\cite{kang2018self,wu2021seq2bubbles}, multi-task learning~\cite{ma2018modeling,yang2022cross}, etc. While our paper mainly focuses on the general setting where only the user-item interactions are assumed as input, the proposed methodology can be trivially extended to other cases to incorporate more information.

\section{Experiments}


\paragraph{Datasets.}
We perform extensive experiments on three publicly accessible datasets from various domains with different sparsities. 
MovieLens~\cite{harper2015movielens} is a widely adopted benchmark dataset for collaborative filtering. We use two versions, namely MovieLens(ML)-100K and MovieLens(ML)-1M. Yelp is a dataset of user ratings on businesses, and we use the filtered subset created by~\cite{he2017fast} for evaluation. Gowalla~\cite{cho2011friendship} is collected from a popular location-based social network, which contains users’ check-in history with time spanning from February 2009 to October 2010.
Each user's interactions are sorted by the timestamps ascendingly. Then the testing data (resp. validation data) comprise the last (resp. second to last) interacted item of each user, while the remaining are used as training data. 
The statistics of filtered datasets are given in Table~\ref{tab:dataset}.

\begin{table}[t]
	\centering
	\scalebox{0.88}{
	\begin{tabular}{c|c|c|c|c}
		\toprule
		Datasets & \#Users & \#Items & \#Interactions & Density  \\
		\midrule
		MovieLens & 6039 & 3415 & 999611 & 4.847\%  \\
		Yelp & 23056 & 15575 & 648687 & 0.181\%  \\
		Gowalla  & 72454 & 56173 & 1360493 & 0.033\%  \\
		\bottomrule
	\end{tabular}}
	\vspace{-5pt}
	\caption{Statistics of three datasets.} \label{tab:dataset}
	\vspace{-10pt}
\end{table}

\paragraph{Evaluation Protocol and Metrics.}
Following~\cite{he2017neural,tay2018latent,rendle2009bpr}, we adopt the \textit{leave one out} protocol for model evaluation. 
We evaluate the ranking performance of the proposed model based on two widely used metrics: \textit{Hit Ratio at K} (HR@$K$), and \textit{Normalized Discounted Cumulative Gain at K} (NDCG@$K$). 

\paragraph{Competitors.}
We consider nine baseline models for collaborative filtering, including four classic collaborative filtering models (PMF~\cite{koren2009matrix}, BPR~\cite{rendle2009bpr}, SVD++~\cite{koren2008factorization}, NeuMF~\cite{he2017neural}), a non-sampling approach ENMF~\cite{chen2020efficient} and two hard negative mining methods (IRGAN~\cite{wang2017irgan}, SD-GAR~\cite{jin2020sampling}).

\subsection{Performance Comparison}
We report experiment results of HDCCF and other comparative models in Tab.~\ref{tab:comparison}. As we can see, the proposed HDCCF consistently outperforms other comparative methods and achieve state-of-the-art results w.r.t. different metrics throughout three dataset. Specifically, HDCCF on average achieves $4.08\%$ improvement for HR@10, $7.61\%$ improvement for NDCG@10 and $6.21\%$ improvement for NDCG@50. The results demonstrate that HDCCF is a powerful approach for recommendation in implicit feedback. There are some other findings. First, four classic methods (i.e., PMF, BPR, SVD++, NeuMF) have the worst performance, which implies the important role of negative mining on promising recommendation performance. Second, SVD++ has better performance than PMF on average since it considers user neighbor information while PMF fails to do so, which also justifies the design of our auxiliary losses. 
Third, HDCCF outperforms adversarial hard negative mining methods, which demonstrate the effectiveness of HDCCF for mining reliable and hard negative instances with more simple and flexible designs.

\subsection{Ablation Study}
We conduct a series of ablation studies to investigate the necessities of some key components in our model and how these components contribute to the overall results.  

\begin{figure}[t]
\centering
\subfigure[ML-100K]{
\label{Fig.ab.1}
\includegraphics[width=0.233\textwidth]{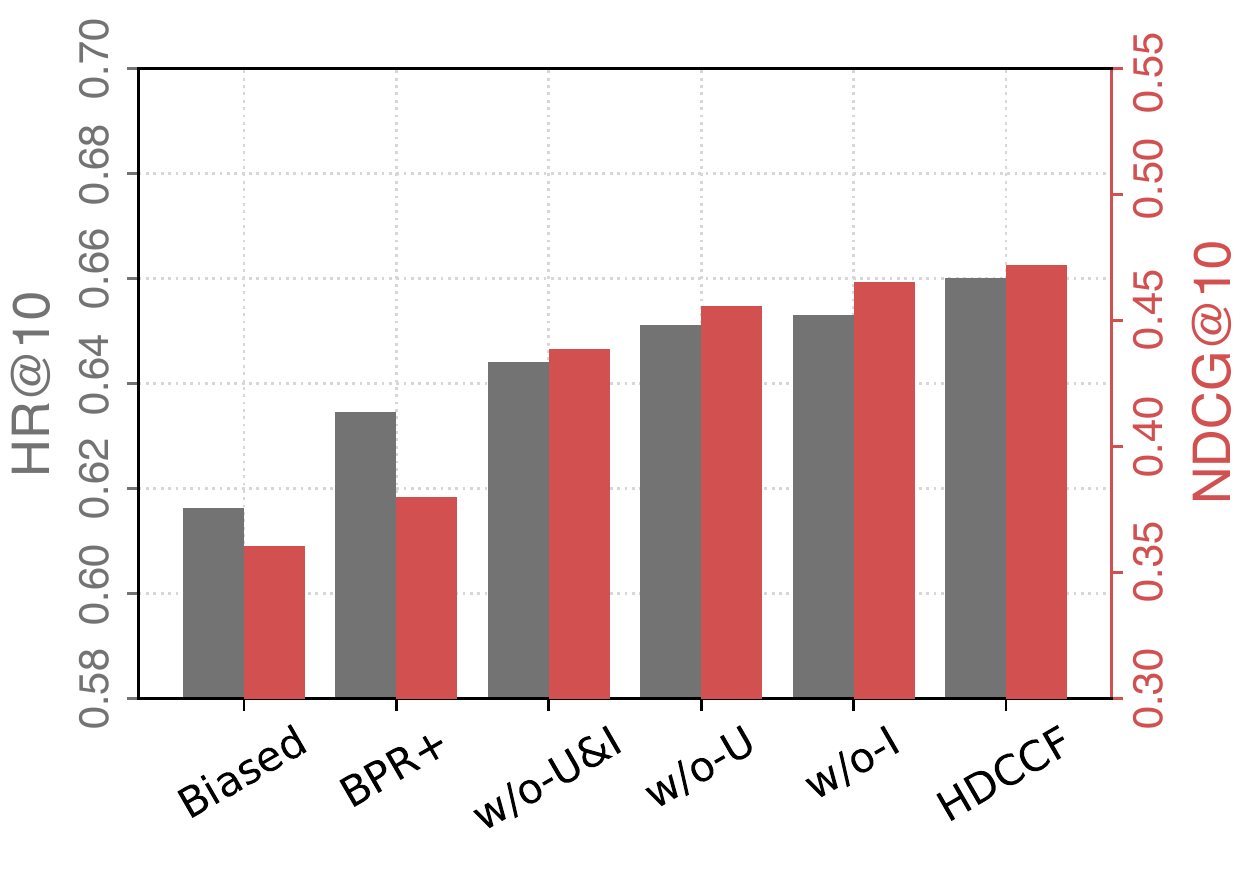}}
\subfigure[ML-1M]{
\label{Fig.ab.2}
\includegraphics[width=0.233\textwidth]{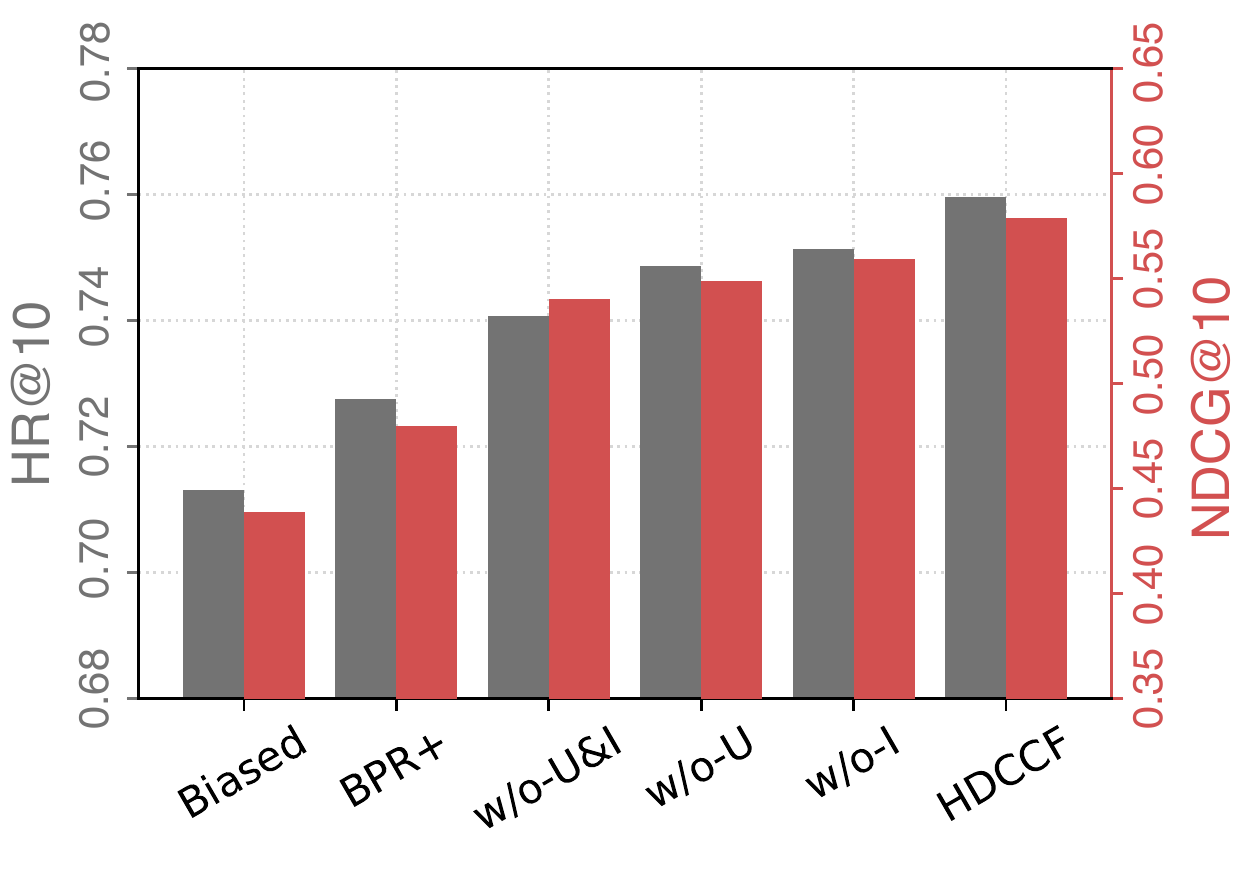}}
\vspace{-5pt}
\caption{Ablation studies on MovieLens for HDCCF.\vspace{-5pt}}
\label{Fig.rq2}
\end{figure}

\paragraph{Debiased Contrastive Losses.}
We compare the performance of HDCCF and five variants of it with simplified loss functions by removing or replacing the contrastive losses. Specifically, we have two findings in Fig.~\ref{Fig.rq2}. First, our HDCCF consistently outperforms the variants of HDCCF which remove $\tilde{\mathcal{L}}_{i-i}$ (w/o-I), $\tilde{\mathcal{L}}_{u-u}$ (w/o-U) and both auxiliary contrastive losses (w/o-U\&I), respectively. This is because $\tilde{\mathcal{L}}_{i-i}$ and $\tilde{\mathcal{L}}_{u-u}$ can take the advantage of hard negative mining to fully exploit the relations of neighbored users and items. Second, the variant of HDCCF that only preserves the user-item contrastive loss (w/o-U\&I) also consistently outperforms the variant that replaces the contrastive loss by BPR loss (BPR+). This result validates that the good recommendation attributes to the proposed loss functions rather than other model designs. Third, HDCCF outperforms the biased variant by a large margin, which demonstrate the importance of removing false negative instances especially in our case where false negatives could be either observed or unobserved. 
 
\begin{figure}[t]
\centering
\subfigure[HDCCF]{
\label{rq4.mask}
\includegraphics[width=0.233\textwidth]{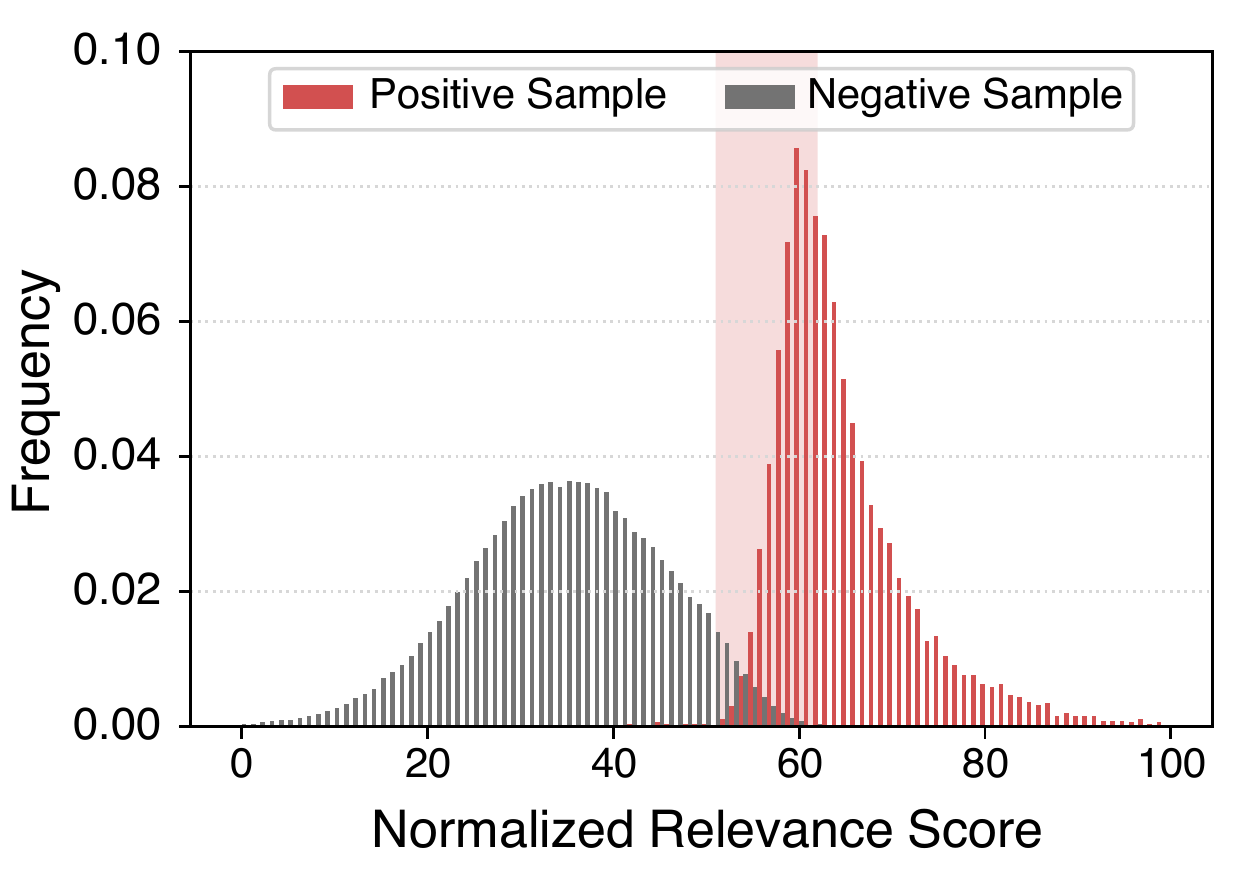}}
\subfigure[Unmodulated variant]{
\label{rq4.dot}
\includegraphics[width=0.233\textwidth]{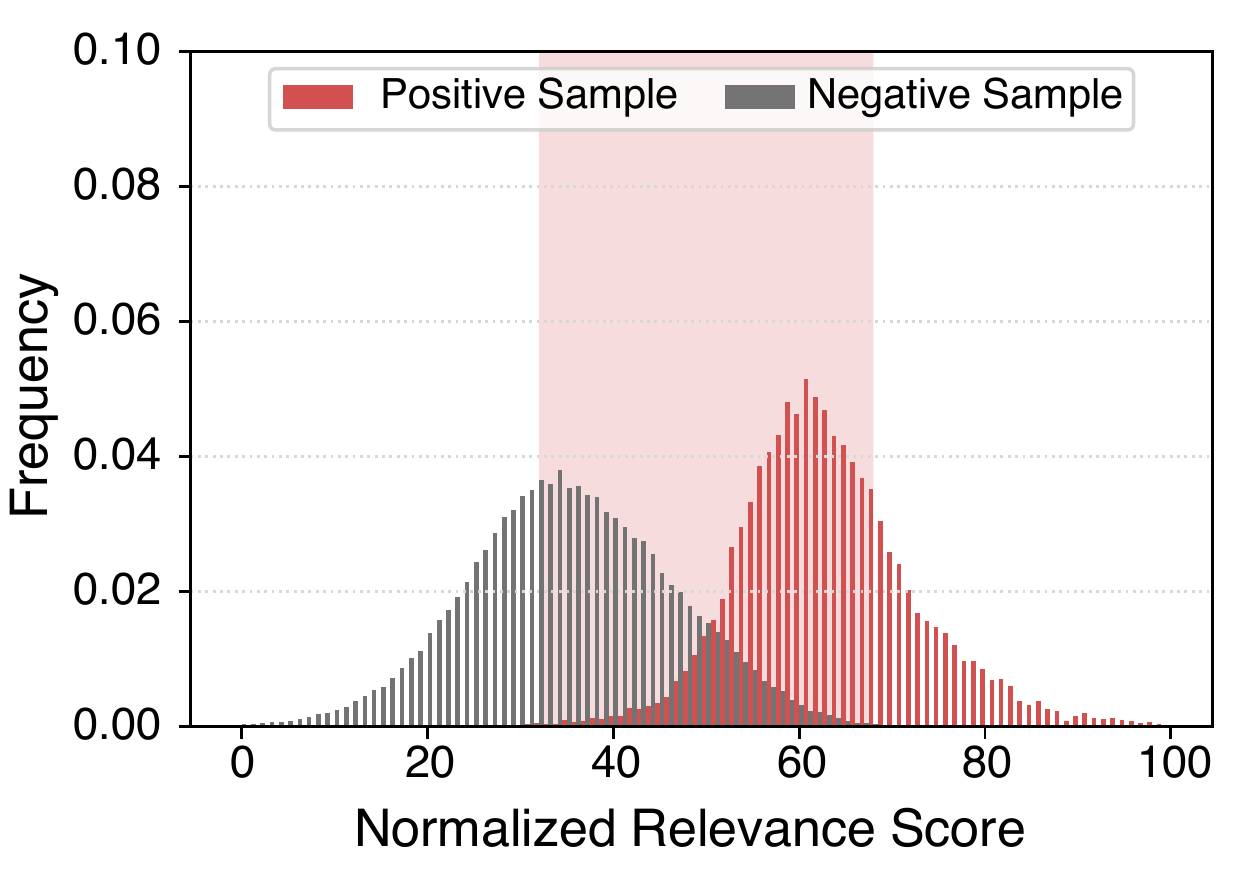}}
\vspace{-5pt}
\caption{Positive and negative user-item relevance score distributions on ML-100K dataset.\vspace{-5pt}}
\label{Fig.rq4}
\end{figure}

\paragraph{Modulated Similarity Model.}
We compare HDCCF with a variant with a fixed modulating vector, i.e., an all-ones modulating vector. Specifically, we visualize the distributions of user-item relevance scores (which are normalized to range $0$ to $100$) for two models on ML-100K dataset in Fig.~\ref{Fig.rq4}. 
As shown in the figure, the modulated similarity model pushes the upper bound of negative instances' relevance scores to the left, and the lower bound of positive instances' relevance scores to the right (which is more significant than the former one). 
Consequently, the intersecting interval (which is filled with red shade in the figure) of our modulated similarity model is considerably smaller than the unmodulated variant. 
This empirical result conforms to the conjecture that the modulation mechanism is helpful to distinguish positive interactions against negative instances by using a more flexible user and item representations.
 

\subsection{Hyper-parameter Analysis}

\begin{figure}[t]
\centering
\subfigure[ML-100K, $\omega_i^+$]{
\label{Fig.hp.1}
\includegraphics[width=0.23\textwidth]{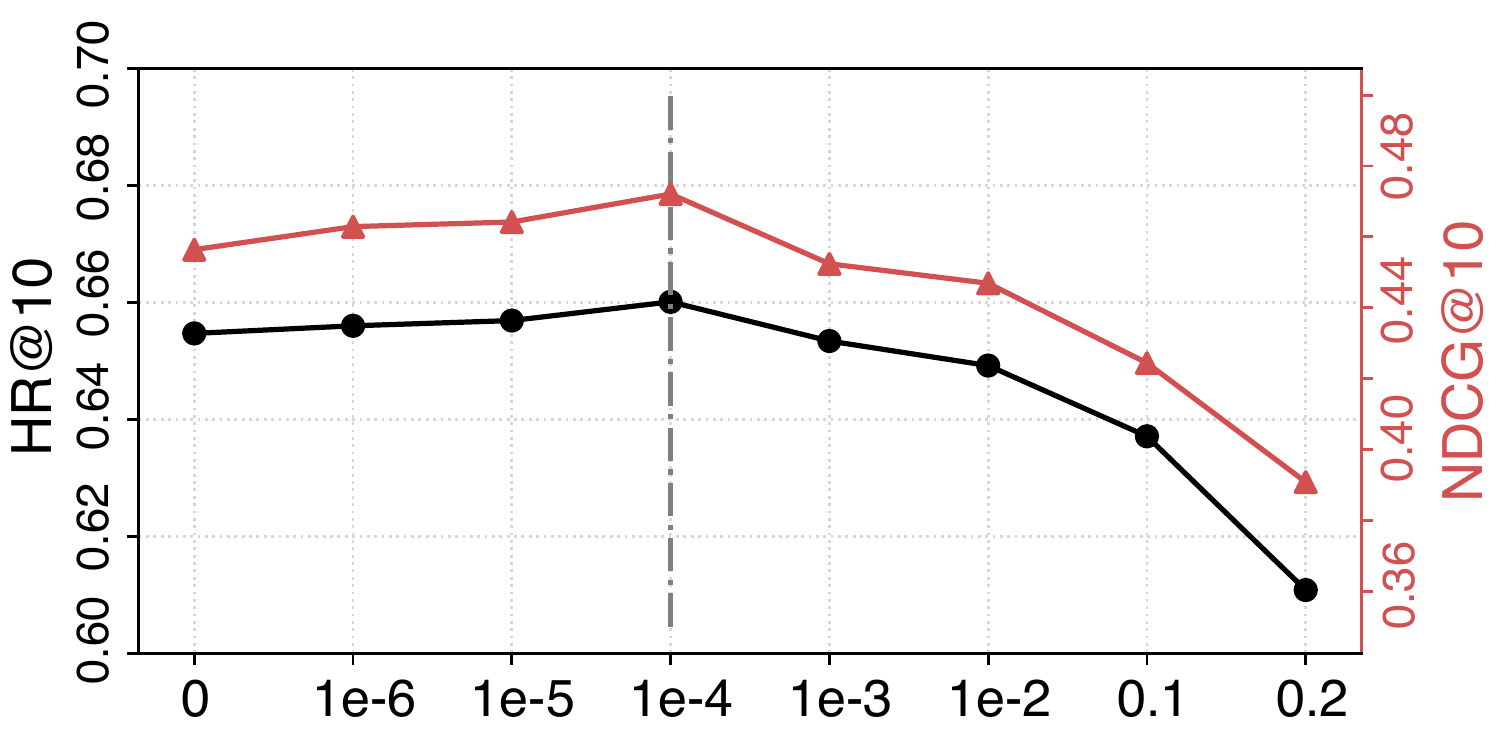}}
\subfigure[ML-1M, $\omega_i^+$]{
\label{Fig.hp.2}
\includegraphics[width=0.23\textwidth]{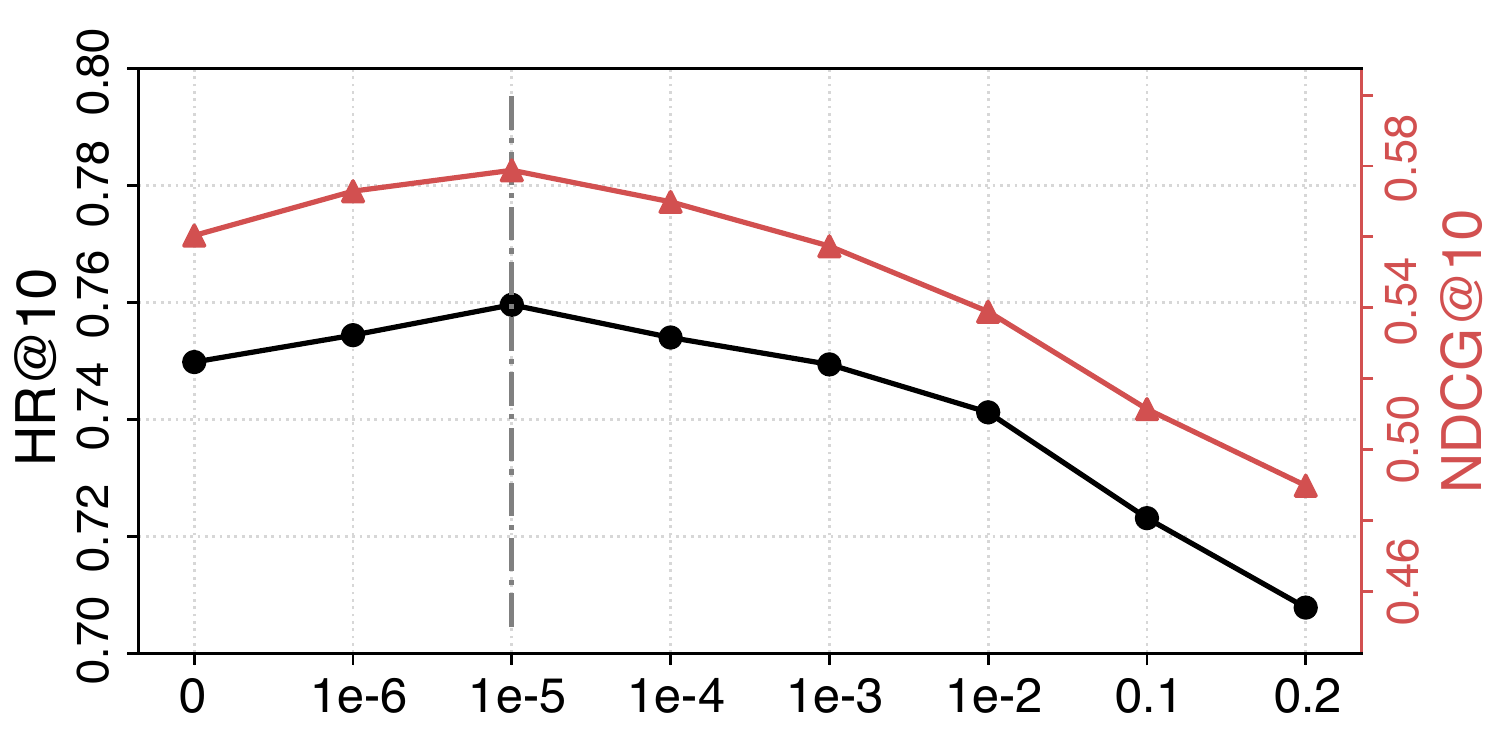}}
\subfigure[ML-100K, $\omega_u^+$]{
\label{Fig.hp.3}
\includegraphics[width=0.23\textwidth]{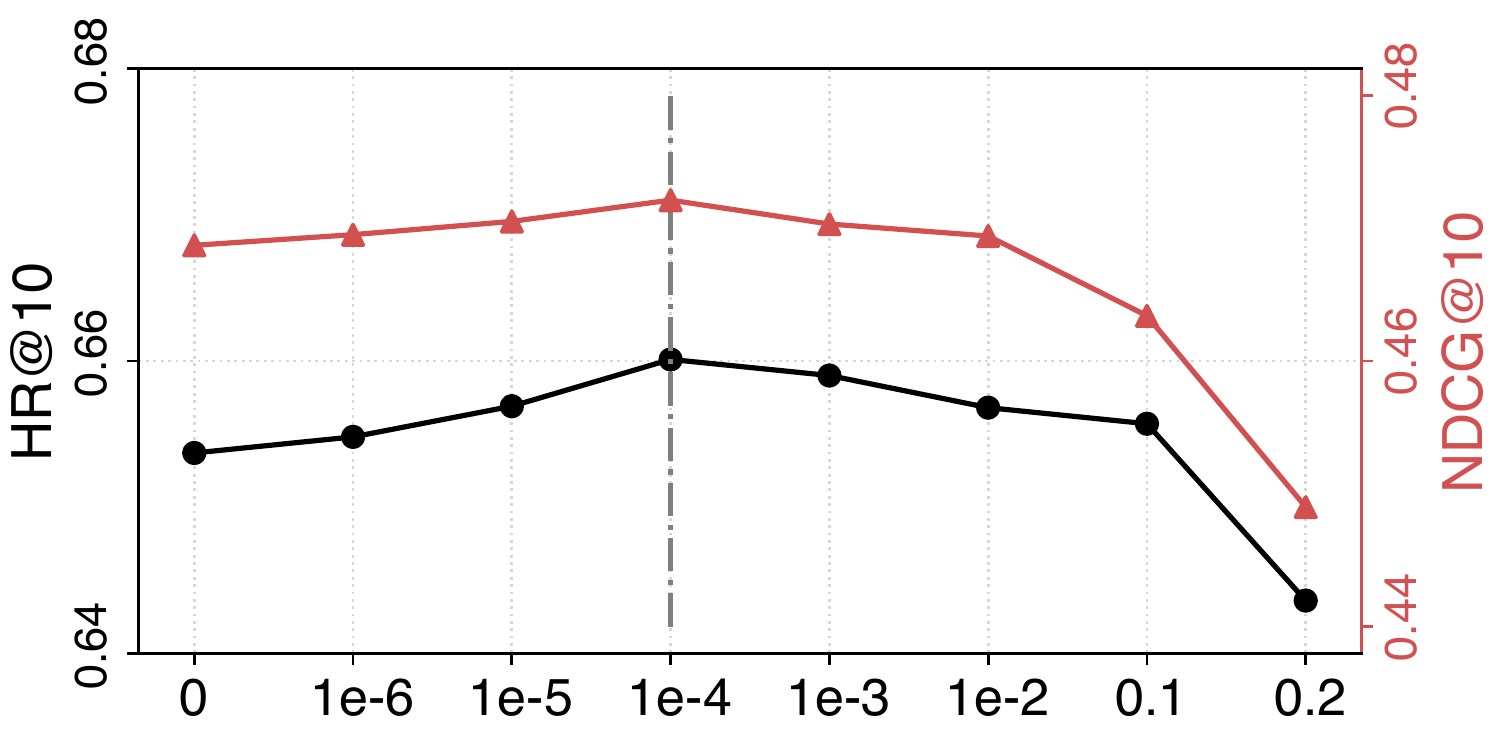}}
\subfigure[ML-1M, $\omega_u^+$]{
\label{Fig.hp.4}
\includegraphics[width=0.23\textwidth]{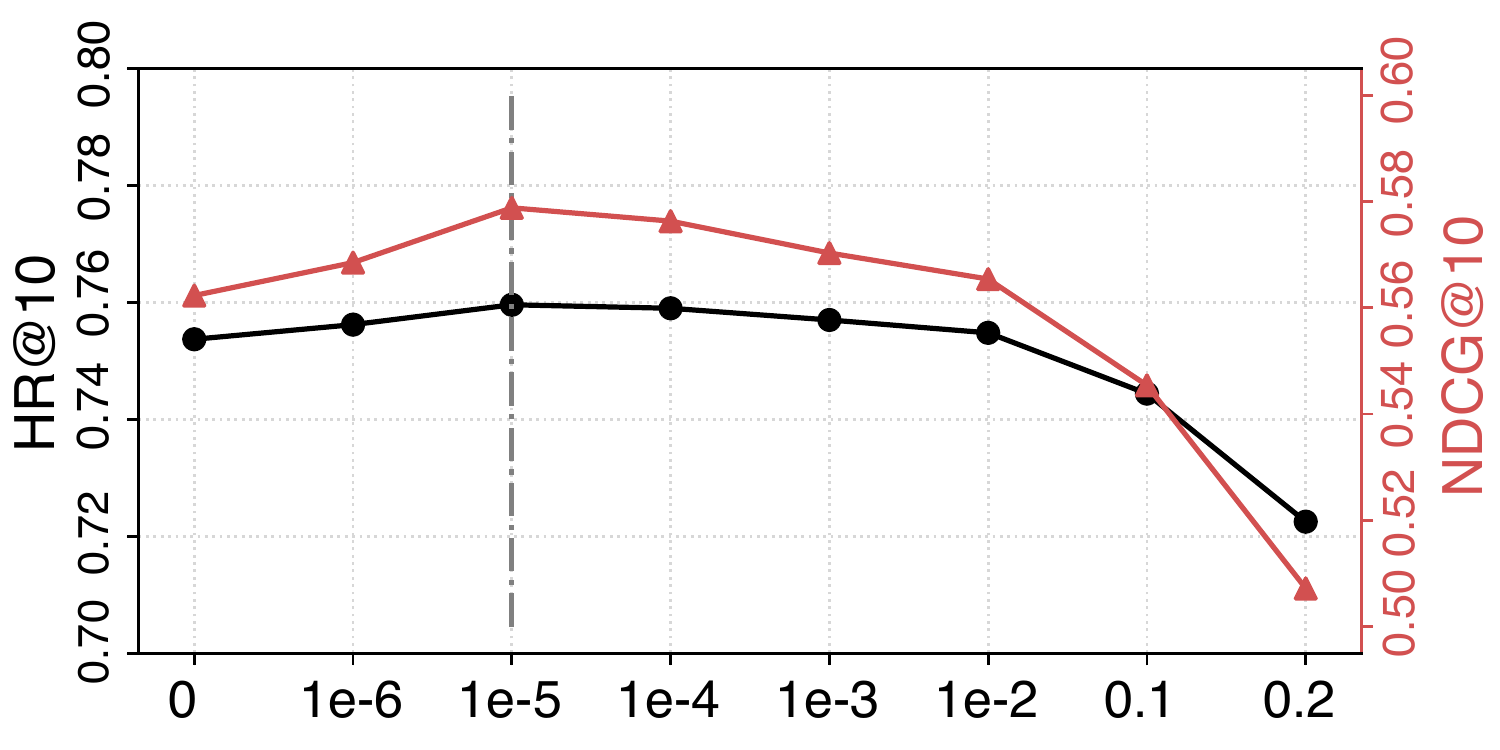}}
\subfigure[ML-100K, $P$]{
\label{Fig.hp.5}
\includegraphics[width=0.23\textwidth]{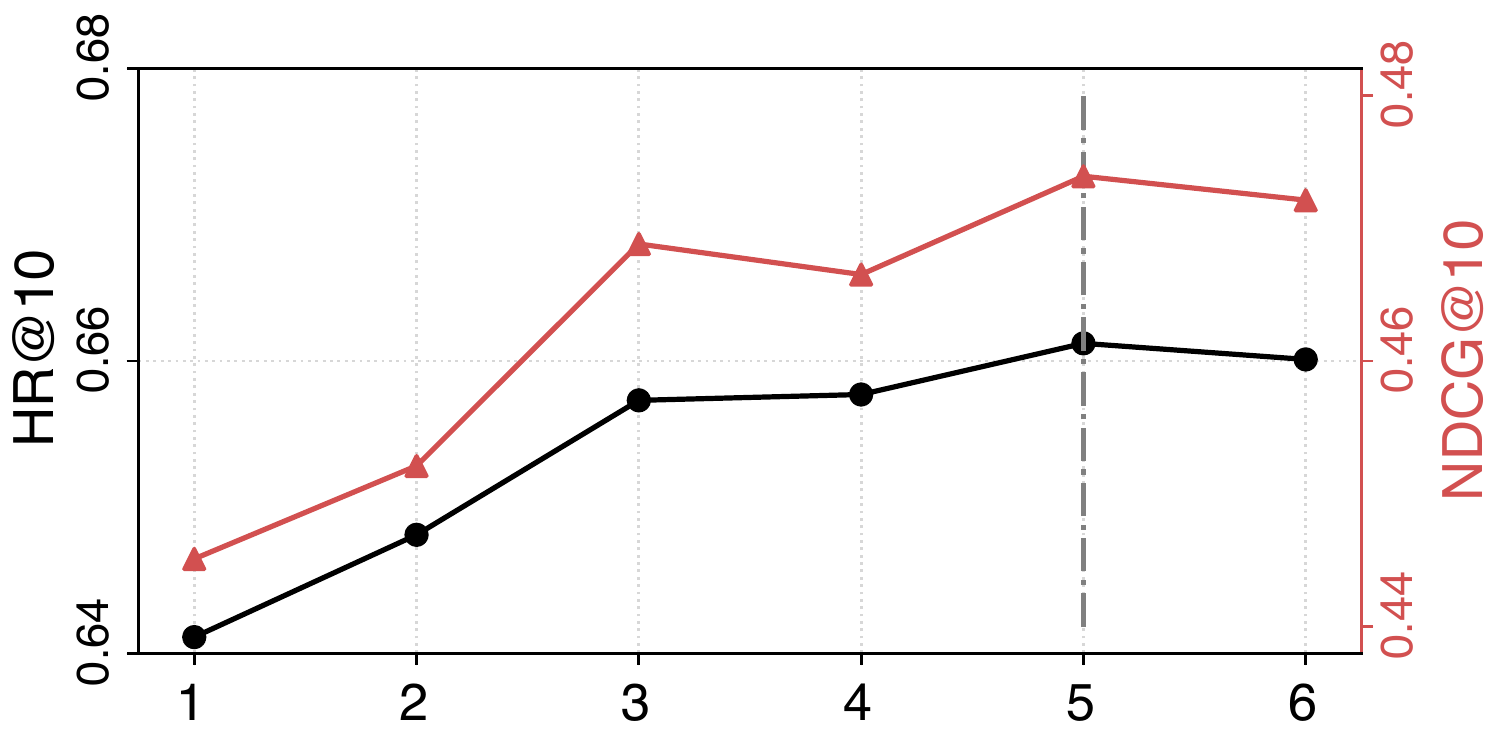}}
\subfigure[ML-1M, $P$]{
\label{Fig.hp.6}
\includegraphics[width=0.23\textwidth]{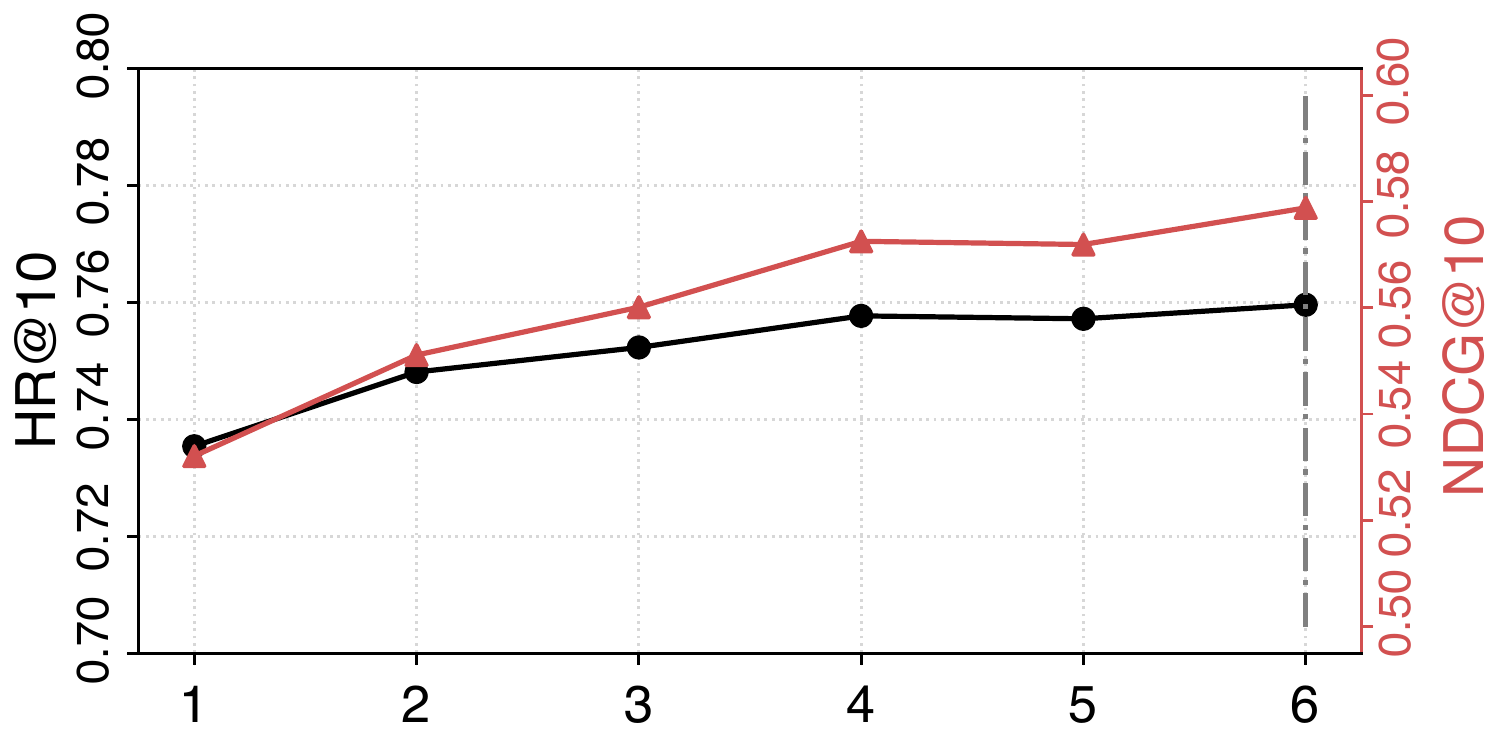}}
\caption{Performance of HDCCF on MovieLens w.r.t. different hyper-parameters. The vertical dotted lines mark the best value.\vspace{-5pt}}
\label{Fig.rq3}
\end{figure}

\paragraph{Impact of Bias Correction Probabilities.}
Two bias correction probabilities $\omega^+_i$ and $\omega^+_u$ are introduced in Sec.\ref{debias} and denote the probabilities that items and users sampled from unobserved data are true negative samples. It is intractable to derive the precise values of $\omega^+_i$ and $\omega^+_u$, and they are different across datasets. Therefore, we treat them as hyper-parameters in implementation and here study how these parameters affect the recommendation performance. Figure~\ref{Fig.rq3}(a)-(d) show HR@10 and NDCG@10 of HDCCF with $\omega^+_i$ and $\omega^+_u$ varying from $0$ to $0.2$ and other hyper-parameters unchanged. As we can see, a certain range of values for $\omega^+_i$ and $\omega^+_u$ can correct the sampling bias, while out-of-range $\omega^+_i$ and $\omega^+_u$ will harm the recommendation performance. 

\paragraph{Impact of Positive Neighbor Sampling Number.}
For our neighbor sampling strategy, $P$ is the hyper-parameter denoting the number of positive user (or item) neighbors for each interaction in the batch. 
Intuitively, larger $P$ brings higher time and space costs, but improves the stability of training process. To validate this conjecture, we study how $P$ affects the recommendation performance. Figure~\ref{Fig.rq3}(e)-(f) show the performance of HDCCF with $P$ varying from 1 to 6 and other hyper-parameters unchanged. As we can see, larger $P$ can bring up improvements for recommendation performance, since it can reduce the variance brought by sampling over observed interactions (which helps to stabilize the training process and alleviate over-fitting) and improve the hard negative mining ability by contrasting with more instances.

\section{Conclusion}
In this paper, we propose hardness-aware debiased contrastive collaborative filtering framework, which resolves the dilemma of hard negative mining and the reliability of negative instances. 
Comprehensive experiments and ablation studies on three real-world datasets demonstrate the effectiveness of our framework.


\small
\bibliographystyle{named}
\bibliography{ijcai22}


\clearpage
\appendix

\section{Proof for Proposition 1} \label{proof-pro1}
 
 \begin{proof}
 For convenience of analysis, we first assume the mini-batch is the whole observation dataset. Then, $\mathbb{E}[n_{i'}^{u-i}]$ is given by:
 \begin{equation}
 \begin{split}
    \mathbb{E}_{(u,i)\sim p^o\atop i^+ \sim p^o_u}\left[n_{i'}^{u-i}\right] &= P\cdot \sum_{(u^*,i^*)\in \mathcal{D}} \mathds{1}[u^*\in \mathcal{U}_{i'}] \cdot \frac{1}{|\mathcal{I}_{u^*}|}\\
    &= P\cdot \sum_{u^*\in \mathcal{U}_{i'}} \frac{\sum_{i^* \in \mathcal{I}_{u^*} }\mathds{1}[u^*\in \mathcal{U}_{i'}]}{|\mathcal{I}_{u^*}|}\\
    &= P\cdot \sum_{u^* \in \mathcal{U}_{i'}} 1\\
    &= P\cdot |\mathcal{U}_{i'}|.
 \end{split}
 \end{equation}
 Since a mini-batch of size $M$ is uniformly sampled from the observation dataset, the final result has a scaling factor $(M-1)/(|\mathcal D|-1)$, i.e.,  $\mathbb{E}[n_{i'}^{u-i}] = P\cdot |\mathcal{U}_{i'}|\cdot (M-1)/(|\mathcal D|-1)$
 \end{proof}

\section{Proof for Proposition 2} \label{proof-pro2}
 \begin{proof}
 For the term $e^{f(u,u^-)}$ in Eqn.~\eqref{eqn:u-u}, we rename it as $e^{f(u_a,u_b)}$ for convenience of analysis. A negative user $u'$ appears in $\mathcal{L}_{u-u}$ when $u' = u_b$. Then, $\mathbb{E}[n_{u'}^{u-u}]$ is given by:
 \begin{equation}
 \begin{split}
    \mathbb{E}_{(u,i)\sim p^o\atop u^+ \sim p^o_i}\left[n_{u'}^{u-u}\right] &= \mathbb{E}[n_{u'=u_b}^{u-u}]\\
    &= P\cdot \sum_{(u^*,i^*)\in \mathcal{D}} \mathds{1}[i^*\in \mathcal{I}_{u'}] \cdot \frac{1}{|\mathcal U_{i^*}|}\\
    &= P\cdot \sum_{i^*\in \mathcal{I}_{u'}} \frac{\sum_{u^* \in \mathcal{U}_{i^*} }\mathds{1}[i^*\in \mathcal{I}_{u'}]}{|\mathcal{U}_{i^*}|}\\
    &= P\cdot \sum_{i^* \in \mathcal{I}_{u'}} 1\\
    &= P\cdot |\mathcal{I}_{u'}|.
 \end{split}
 \end{equation}
 The final result also has a scaling factor $(M-1)/(|\mathcal D|-1)$ in consideration of the batch size $M$.
 \end{proof}



\section{Proof for Debiased Contrastive Losses} \label{proof-debias}
\begin{proof}
Assume a observed interaction $(u,i)$ is sampled from $\mathcal{D}$ with distribution $p^o(u,i)$. Given an item $i$, a negative user $u^-$ is sampled from unobserved user set $\mathcal{U}'_i$ with distribution $p_i(u^-)$. The user $u^-$ sampled in this way could be real negative or false negative. Assume $p^+_i(u^-)$ is the probability of observing $u^-$ as a false negative example and $p^-_i(u^-)$ the probability of a real negative example. The ideal loss to optimize for $\mathcal{L}_{u-i}$ should be:
\begin{equation}
\begin{split}
    &\mathcal{L}_{u-i}^{ideal} = -\mathbb{E}_{(u,i)\sim p^o\atop u^- \sim p_i^-, i^- \sim p_u^-}\\ 
    &\left[ \frac{e^{f(u,i)}}{e^{f(u,i)} + \frac{Q}{N}\sum_{k=1}^N e^{f(u_k^-, i)} + \frac{Q}{N}\sum_{k=1}^N e^{f(u, i_k^-)}} \right],
\end{split}
\end{equation}
where $Q$ is a weighting parameter for the analysis. For fixed $Q$ and $N\rightarrow \infty$, it holds that:
\begin{equation} \label{eqn:ideal}
\begin{split}
    &\mathcal{L}_{u-i}^{ideal} = -\mathop{\mathbb{E}}\limits_{(u,i)\sim p^o} \\
    &\left[ \frac{e^{f(u,i)}}{e^{f(u,i)} + Q\mathop{\mathbb{E}}\limits_{u^- \sim p_i^-}[e^{f(u^-, i)}] + Q\mathop{\mathbb{E}}\limits_{i^- \sim p_u^-}[e^{f(u, i^-)}]} \right].
\end{split}
\end{equation}
Now, we focus on the second term in the denominator of Eqn.~\eqref{eqn:ideal}. Suppose $u'$ is a unobserved user for item $i$, the data distribution can be decomposed as $p_i(u') = \omega_u^+ \cdot p_i^+(u') + \omega_u^- \cdot p_i^-(u')$. Therefore, we could write $p_i^-(u')$ as:
\begin{equation} \label{eqn:decompose}
p_i^-(u')=\left\{
\begin{aligned}
&0  &,  (u',i)\in \mathcal{D}, \\
\frac{p_i(u')}{\omega_u^-} &- \frac{\omega_u^+ \cdot p_i^+(u')}{\omega_u^-}  &,  (u',i)\notin \mathcal{D}.
\end{aligned}
\right.
\end{equation}
According to Eqn.~\eqref{eqn:decompose}, the second term in the denominator of Eqn.~\eqref{eqn:ideal} could be rewrite as:
\begin{equation}
\begin{split}
        &F^u(u, i)^{ideal} \\
        &= Q\mathop{\mathbb{E}}\limits_{u^- \sim p_i^-}[e^{f(u^-, i)}]\\ 
        &= Q\left(\frac{1}{\omega_u^-}\mathop{\mathbb{E}}\limits_{u^- \sim p_i}[e^{f(u^-, i)}] - \frac{\omega_u^+}{\omega_u^-}\mathop{\mathbb{E}}\limits_{u^+ \sim p^+_i}[e^{f(u^+, i)}] \right).
\end{split}
\end{equation}
Note that both $u^-$ and $u^+$ are sampled from unobserved user set $\mathcal{U}'_i$. Given the negative sample set $\mathcal{N}_i^-$ which consists of both observed users and unobserved users, we decompose it as $\mathcal{N}_i^- = \hat{\mathcal{N}}_i^- \cup \check{\mathcal{N}}_i^-$, where $\hat{\mathcal{N}}_i^-$ is the observed subset and $\check{\mathcal{N}}_i^-$ is the unobserved subset. The empirical estimate of $F^u(u, i)^{ideal}$ is given by:
\begin{equation}
\begin{split}
        &\tilde{F}^u(\mathcal{N}_i^-, \mathcal{N}_i^+) \\
        &= \frac{Q}{\omega_u^-}\Bigg(\frac{\sum_{u^-\in \check{\mathcal{N}}_i^-}e^{f(u^-, i)}}{|\check{\mathcal{N}}_i^-|} - \frac{\omega_u^+}{ |\mathcal{N}_i^+|+1}\sum_{u^+\in \mathcal{N}_i^+ \cup \{u\}}e^{f(u^+, i)} \Bigg)\\
        &= \sum\limits_{u^-\in \mathcal{N}_i^-} \pi^u_0(u^-,i)\cdot e^{f(u^-, i)} - \sum\limits_{u^+\in \mathcal{N}_i^+ \cup \{u\}} \pi^u_1(u^+,i)\cdot e^{f(u^+, i)},
\end{split}
\end{equation}
where $\pi^u_0(u^-,i), \pi^u_1(u^+,i) \in \mathbb{R}$ are constants w.r.t. $|\mathcal{N}_i^-|$ and $|\mathcal{N}_i^+|$: 
\begin{equation}
\begin{split}
        &\pi^u_0(u^-,i) = \frac{Q\cdot \mathds{1}[(u^-,i)\notin\mathcal{D})] }{\omega_u^-\sum\limits_{u'\in \mathcal{N}_i^-}\mathds{1}[(u',i)\notin\mathcal{D})] }, \\
        &\pi^u_1(u^+,i) = \frac{Q\cdot \omega_u^+ }{\omega_u^-(|\mathcal{N}_i^+|+1) }.
\end{split}
\end{equation}
Note that here we assume $u^+$ sampled from $p_i^+$ is equivalent to $u^+$ sampled from observed interactions, which is reasonable since they are both trustful positive users for $i$. In practice, we let $Q=|\mathcal{N}_i^-|$ for simplicity. The empirical estimate of the third term in the denominator of Eqn.~\eqref{eqn:ideal} could be derived in the same way by exchanging symbols $u$ and $i$. 

For $\mathcal{L}_{u-u}$ (resp. $\mathcal{L}_{i-i}$), we assume sampling from unobserved user set is equivalent to sampling from real negative user set.
The reason is that positive user neighbors $(u,u^+)$ generally share a number of commonly interacted items, and one can observe them if partial shared items are observed in $\mathcal{D}$. Therefore, unlike user-item pairs, most positive user neighbors $(u,u^+)$ are observed. Then, the unbiased user-user contrastive loss is 
\begin{equation}
\begin{split}
    \tilde{\mathcal{L}}_{u-u} = -&\sum_{(u,i)\in \mathcal{D}}\sum_{u^+ \in \mathcal{N}_i^+} \\
    &\log \frac{e^{f(u,u^+)}}{e^{f(u,u^+)} + \sum\limits_{u^- \in \mathcal{N}_i^-}\pi^{u-u}(u,u^-)\cdot e^{f(u,u^-)}},\\
\end{split}
\end{equation}
where
\begin{equation}
    \pi^{u-u}(u,u^-) = \frac{Q\cdot \mathds{1}[(u,u^-)\notin\mathcal{D}^{u-u})]}{\sum\limits_{u' \in \mathcal{N}_i^-} \mathds{1}[(u',u^-)\notin\mathcal{D}^{u-u})]}.
\end{equation}
The unbiased item-item contrastive loss $\tilde{\mathcal{L}}_{u-u}$ could be derived in the same way by exchanging symbols $u$ and $i$.

\end{proof}

\end{document}